\documentstyle[prl,multicol,aps]{revtex} 
\input epsf

\begin{document}
\draft 
\title{Molecular dynamics simulation of the fragile glass former 
        ortho-terphenyl: a flexible molecule model. II. Collective Dynamics}

\author{S.~Mossa$^{1,2}$, G.~Ruocco$^{2}$ and M.~Sampoli$^{3}$}

\address{$^1$ Center for Polymer Studies and Department of Physics,
Boston University, Boston, Massachusetts 02215}

\address{$^2$ Dipartimento di Fisica and Istituto Nazionale di Fisica per la 
Materia, Universit\`a di Roma La Sapienza, \\
P.le Aldo Moro 2, I-00185, Roma, Italy}

\address{$^3$ Dipartimento di Energetica and INFM, Universit\`a di Firenze,
         Via Santa Marta 3 , Firenze, I-50139, Italy}

\date{\today}
\maketitle
\begin{abstract}
We present a Molecular Dynamics study of the collective dynamics of a 
model for the fragile glass-former orthoterphenyl. In this model, 
introduced by Mossa, Di Leonardo, Ruocco and Sampoli 
[S.~Mossa, R.~Di Leonardo, G.~Ruocco and M.~Sampoli,
Phys. Rev. E {\bf 62}, 612 (2000)], the intramolecular interaction
among the three rigid phenyl rings is described 
by a set of force constants whose value has been fixed in order to obtain
a realistic isolated molecule spectrum. The interaction between different
molecules is described by a Lennard Jones site-site potential. 
We study the behavior
of the coherent scattering functions $F_t(q,t)$, considering
the density fluctuations of both molecular and phenyl rings
centers of mass; moreover we directly simulate the 
neutron scattering spectra taking into account both the 
contributions due to carbon and hydrogens atoms. 
We compare our results with the main predictions 
of the Mode Coupling Theory
and with the available coherent neutron scattering experimental data. 
\end{abstract}


\begin{multicols}{2}
\section{INTRODUCTION}
\label{introduction}

The understanding of the 
supercooled liquid and glassy phases in molecular systems
is, nowadays, one of the main tasks of the physics of disordered materials
(see~\cite{angell_first,angell} and references therein for a general
review). Two main theories provide us a description of the 
glass transition respectively from a thermodynamical and
dynamical point of view.

The first one (see~\cite{mezpar} and references therein)
is based on first-principles computation of the {\em equilibrium}
thermodynamics of glasses and consider the glass transition as
a true thermodynamic transition. In this context,
the incoming of the glassy state is associated with an entropy crisis,
i. e. the vanishing of the {\em configurational} entropy 
of the thermodynamically relevant states. 
 
The second approach is the Mode Coupling Theory (MCT)~\cite{mct1,mct2}
which studies the long-time structural dynamics 
and its relation with the glass transition; 
in this context this transition has to be considered
not as a regular thermodynamical phase transition involving singularities
of some observables, but as a kinetically induced transition
from an ergodic to a non-ergodic behavior. The structural dynamics
becomes so ``slow'' that the system of interest appeared
frozen on the experimental time scales.
  
On the experimental side, the collective dynamics of 
a huge variety of molecular systems 
has been investigated 
by means of several experimental techniques;
colloids~\cite{colloids1,colloids2} and ortho-terphenyl 
(OTP)~\cite{hwang,bartfujleg,tollschowutt,tollwutt} are 
the most widely studied {\em fragile}~\cite{angell_first} 
supercooled liquids. They are only examples of an enormous 
experimental work (the interested reader is referred to
Ref.~\cite{angell} for an accurate comprehensive review).
Moreover, in the last ten years the analysis of experimental results
has been flanked  by extensive use of numerical techniques,
mainly Molecular Dynamics (MD) and Monte Carlo (MC) simulations.
The almost exponential growth of computational 
capabilities allows us to reach simulation times of the order 
of microseconds for simple monoatomics systems and of
several tens nanosecond in the case of complex molecular systems; 
such performances definitely permit
comparisons of numerical results with the structural very long time 
properties of the real systems. 
We remember, among many others, the results concerning
the structural dynamics of two molecular liquids like
SPC/E water~\cite{water_coll1}
and OTP~\cite {lewa,scioradel}.
We emphasize here the fact that the models involved in these studies
are {\em molecular} and {\em rigid} in the sense that 
they have a structure and take into account
the orientational properties of the molecules but they 
disregard the role played by internal degrees of freedom
on the overall dynamical behavior of the system. 

In a recent paper~\cite{modilruosam}
we have introduced a new model for the intramolecular dynamics of 
orthoterphenyl (OTP)
($T_m=329$ K, $T_c\simeq 290$ K, $T_g=243$ K), one of the most deeply
studied substances in the liquid, supercooled and glassy state.
The introduction of such a flexible model allows 
us to understand the role of internal degrees of freedom
in the short time ({\em fast}) dynamics~\cite{monfiomasc} 
taking place on the time scale of a few picoseconds and to study 
their possible coupling with the long time center of mass dynamics~\cite{fastmd}.
Moreover would allow us to emphasize once more
the universality of the Mode Coupling Theory (MCT)
approach for supercooled liquids and, in particular, of its molecular version 
in the case of complex molecular liquids.
In Ref.~\cite{modilruosam} we introduced in details
the intermolecular model and we showed
some results based on MD calculations;
the self (one particle) dynamical properties
of a system composed of 108 molecules 
have been studied in details
and we have found good agreement with the main predictions of the MCT
and with the experimental results related to the self intermediate
scattering function~\cite{petbarpuj,kiebbardeb}
and the self diffusion~\cite{calldougfal,fujgeilsill}. 
In the present paper we complete the picture 
considering the collective highly cooperative 
structural dynamics controlling
the rearrangement of big portions of the system.
We describe the dynamics of the molecules at the level of
molecular centers of mass and at the level of 
phenyl rings center of mass.
Moreover we calculate the neutron coherent 
scattering function taking into account both the contributions
due to carbon and hydrogen atoms in order to make a direct comparison
with the experimental results.

The paper is organized as follows:  
in Sec.~\ref{MCT} we summarize the main predictions
of the Mode Coupling Theory (MCT) and we give a schematic 
introduction to the Molecular Mode Coupling Theory (MMCT); 
in Sec.~\ref{model} we briefly describe the model and we
recall some computational details.
In Sec.~\ref{mol_rings}
we study the temperature and momentum dependence of the
collective dynamics
of the molecular and of the phenyl rings centers of mass.
In Sec.~\ref{neutrons} we make a deep comparison
between the experimental results of neutron scattering
and the simulated neutron spectra calculated taking into account
the interactions with both carbon and hydrogen atoms. 
Finally, Sec.~\ref{conclusions} contains a discussion of the
results obtained and some conclusions.
 
\section{MODE-COUPLING AND MOLECULAR MODE-COUPLING THEORY}
\label{MCT}

As we have already stressed in Sec.~\ref{introduction} two 
main theories, one intrinsically thermodynamical~\cite{mezpar}
and the other purely dynamical~\cite{mct1,mct2} have 
been developed up to now to describe the phenomenology
observed in the glass transition. In this paper we 
will make a comparison of our numerical results with the
main predictions of MCT about the center of mass structural relaxation 
dynamics.
The reason for this is twofold: first of all this is, indeed, the theory 
taking into account the states actually
accessible by the system on the time scale of a typical simulation. 
Moreover, even if some experimental results, like the presence of the so-called 
knee characterizing the low-frequency behavior of the light scattering
susceptibility~\cite{nomct1,nomct2} or the presence of a cusp
in the non-ergodicity parameter~\cite{nomct3}, seem to contradict 
some of its predictions,
such theory has been verified to hold
for different experimental data on a very wide time region. 
Finally, in the last two years successful
efforts have been made in order to generalize the theory
to liquids of {\em rigid} molecules of {\em arbitrary shape}~\cite{mmct}
(see also Ref.~\cite{mmct2} for the particular case 
of a liquid of {\em linear} rigid molecules)
taking into account both translational and rotational dynamics.

In Ref.~\cite{modilruosam} we have summarized
the main predictions of the so called {\it ideal} MCT
so that here we recall only the 
fundamental equations of the theory,
concerning the {\em collective intermediate scattering function}
and that we will use in the following sections.
MCT interprets the glass formation as a {\it dynamical transition} 
from an ergodic to a non-ergodic behavior at a 
cross-over temperature $T_c$; the theory is written 
as a self-consistent dynamical treatment~\cite{mct1} 
of the intermediate scattering function
, i. e. the time correlation of the density fluctuations
of momentum $q$.

This theoretical scheme can be considered the mathematical
description of the physical picture of the {\em cage effect}. 
Following the dynamics of a tagged particle
it is possible to recognize two main dynamical regions.
On small time scale of order of some picosecond,  
(the $\beta$-region) the dynamics of the particle is confined 
into a limited region (the cage) builded up
by the nearest neighbors. In this regime it is possible
to write the intermediate scattering function as
\begin{equation}
F(q,t)=f(q)+h(q) \sqrt{\frac{\vert T-T_c\vert}{T_c}} G_{\pm}(t/\tau_{\sigma})
\label{betaregion}
\end{equation}
where $f(q)$ is the {\it non-ergodicity parameter}
(also referred to as the {\it Debye-Waller factor}), $h(q)$ is an amplitude
independent of temperature and time and the $\pm$ in $G_{\pm}$
corresponds to time larger or smaller with respect to $\tau_\sigma$,
a parameter that fix the time scale of the $\beta$-process.
At this stage, the time dependence of the correlation functions
is all embedded in the {\it q-independent} function  $G_{\pm}$.
$G_{\pm}(t)$ is asymptotically expressed by two power laws,
respectively the {\it critical decay} 
\begin{eqnarray}
G_{+}\left(t/\tau_{\sigma}\right)=
(t/\tau_{\sigma})^{-a}&\;\;\;\;& \mbox{$\tau_0<<t<<\tau_\sigma$}
\label{apower}
\end{eqnarray}
and the {\it von Schweidler law},
\begin{eqnarray}
G_{-}\left(t/\tau_{\sigma}\right)=
-(t/\tau_{\sigma})^b&\;\;&\;\; \mbox{$\tau_\sigma<<t<<\tau_\alpha$};
\label{bpower}
\end{eqnarray}
characterized by the {\it temperature and momentum independent} 
exponents $a$ and $b$; here $\tau_\alpha$ 
is the {\em structural relaxation time}.

At time scales of order of $\tau_\alpha$
the cages start to break down and the particle starts
to diffuse approaching pure brownian motion. This long time part of the
dynamics is the so called $\alpha$-region and is well described by a
stretched exponential function
\begin{equation}
F(q,t)\simeq f(q)\,\exp\left\{-\left(\frac{t}{\tau_\alpha}\right)
^{\beta_\alpha}\right\}
\label{stretched}
\end{equation}
which verifies the {\em time-temperature superposition principle}
(TTSP).
The $\alpha$ time scale $\tau_\alpha$ depends on 
temperature trough a power law of the form
\begin{equation}
\tau_\alpha\propto(T-T_c)^{-\gamma}
\label{powerlaw}
\end{equation}

The momentum dependence of the dynamical parameters 
in the collective dynamics case 
is not trivial as in the single-particle case. 
When looking at the structural collective dynamics
studying different values of the momentum $q$ means
studying the highly cooperative time evolution of cages 
of average dimension $2\pi/q$; 
it is clear that such time evolution is strongly coupled to
the static topological structure of the system.
More precisely, MCT predicts that the parameters $f(q)$, 
$h^{-1}(q)$, $\beta_{\alpha}$ and  $\tau_\alpha(q)$
oscillate {\em in phase} with the static structure factor $S(q)$.

The result concerning the momentum dependence of the 
collective relaxation time is quite general
and is well known as {\em de Gennes narrowing}~\cite{narrowing}.
A general relation~\cite{madden} holds among the one-particle
$\tau_s$ and the collective $\tau_c$ relaxation times, namely
$\tau_c(q)\simeq S(q)\tau_s(q)$; if the diffusion limit is appropriate
for $F_s(q,t)$, i.e. for values of $q$
close to the first peak of the static structure factor,
we obtain 
\begin{equation}
\tau_c(q,T)\simeq\frac{1}{D(T)}\frac{S(q)}{q^2}
\label{tau_resc}
\end{equation}
where $D(T)$ is the diffusion coefficient at temperature $T$
and $S(q)$ is supposed to be nearly temperature independent.

In summary,
the momentum dependence of the collective dynamics is non trivially
driven by the static structure of the system.
In particular, for {\em molecular} systems, the small length scale structure is 
determined by orientational properties of the single molecules;  
a pure molecular translational dynamics
will, obviously, loose all the dynamical features
controlled by the high momentum part of the static structure factor.

An important step toward a correct explanation of the dynamics
of molecular systems is to write down 
a Molecular Mode Coupling Theory (MMCT)~\cite{mmct,mmct2} taking into account 
both translational and rotational degrees of freedom.
If we consider a system of $N$ identical {\em rigid} molecules
of arbitrary shape described by their center of mass positions
$\bar r_j(t)$ and by the Euler angles
$\Omega_j(t)=(\phi_j(t),\theta_j(t),
\chi_j(t))$ we can write the time-dependent microscopic one-particle density as
\begin{equation}
\rho(\bar r,\Omega,t)=\sum_{n=1}^{N} \delta [\bar r - \bar r_n
(t) ]\delta [\Omega,\Omega_n(t)].
\end{equation}
Expanding with respect to the complete set of functions
given by the plane waves and the Wigner matrices $D^l_{mn}(\Omega)$,
we have the {\em tensorial} density modes
\begin{equation}
\rho_{lmn}(\bar q,t)=i^l(2l+1)^{1/2}\sum_{n=1}^{N}e^{i\bar q\cdot \bar r_n(t)}
D^{l*}_{mn}\left[\Omega_n(t)\right].
\end{equation}
Then, the generalization of the intermediate scattering function  
to the molecular case is the tensorial quantity
\begin{equation}
S_{lmn;l'm'n'}(\bar q,t)=\frac{1}{N}
\langle \delta\rho^*_{lmn}(\bar q,t)\delta\rho^*_{l'm'n'}(\bar q)\rangle.
\end{equation}
These correlators are directly related to experimental quantities~\cite{mmct2};
for $l=l'=0$, they describe the dynamics of translational degrees of
freedom which can be measured by neutron scattering
when looking at the center of mass low frequency part of the spectrum; 
if the molecules possess a permanent dipolar moment, the correlators with $l=l'=1$ give
information related to dielectric measurements and $l=l'=2$ is finally related
to the orientational contribution to light scattering.
At this stage, provided the static angular correlators
$S_{lmn;l'm'n'}(\bar q,0)$ and the number density,
it is possible to give a closed set of equations
for the matrix $\mathbf S$ that completely 
solve the problem of a liquid of rigid molecules.

The problem is that MMCT seems to be not enough 
for a high structured molecular system like OTP;
we will see that it is not possible 
to explain some features of the momentum dependence
of the structural dynamics without taking
into account the internal degrees of freedom 
(i. e. rotations of the side rings with respect to the central one)
that turn out to be strongly coupled to 
the long time behavior of the density fluctuations. 
   
\section{MODEL AND COMPUTATIONAL DETAILS}
\label{model}

In this section we give a brief description of the model
and we address the reader to Ref.~\cite{modilruosam} for details.

In our model, the OTP molecule is constituted by
three rigid hexagonal rings of side $L_a=0.139$ nm 
representing the phenyl rings; two adjacent vertices
of the central ring are bonded to one vertex of the two lateral rings by
bonds of equilibrium length $L_b=0.15$ nm. In the isolated
molecule equilibrium position, the two lateral rings lie
in planes that form an angle of about $54^{o}$ with respect
to the central ring's plane.
In the model the two lateral rings are free to rotate along the molecular bonds,
to stretch along the bonds and to tilt out of the plane identified by the
central ring. The intramolecular potential is then written as 
a sum of harmonic and anharmonic terms each one controlling
one of these features. Every term is multiplied by a coupling constant
whose actual value is determined in order to have a realistic isolated  
molecule vibrational spectrum.
The intermolecular interaction is of the site-site Lennard-Jones type;
each of these sites corresponds to a vertex of a
hexagon and is occupied by a fictious atom of mass $M_{CH}=13$ amu
representing a carbon-hydrogen pair. 
The actual values of the parameters $\sigma_{LJ}$ and $\epsilon_{LJ}$
have been fixed in order to have the first maximum of the
static structure factor $S(q)$ in the experimentally 
determined position~\cite{barbertchi}
and to obtain the correct diffusional
properties~\cite{calldougfal,fujgeilsill}; 
the cut-off has been fixed to the value $r_c=1.6$ nm$^{-1}$.
It is worth noting here that obviously
the parameterization of the potential cannot be perfect;
in our case it is possible to reproduce quite well 
the experimental results on the whole investigated temperature range
shifting the MD thermodynamical points at temperatures of $20$ K
above their true values.  
The MD simulated system is composed of $108$ molecules ($324$ phenyl rings
for a total of $1944$ Lennard Jones interaction sites); at each time step the
intramolecular and intermolecular interaction forces are calculated
and the equation of motion for the rings are solved for the
translational and rotational part separately. 

Wide temperature and momentum ranges have been 
investigated for values of temperature
$380 \leq T\leq 440 K$ and momentum $2\leq q\leq 30$ nm$^{-1}$  
(the runs details are shown in Table III
of Ref.~\cite{modilruosam}) and the total simulation time
is of almost one hundred nanoseconds.

\section{MOLECULES AND PHENYL RINGS}
\label{mol_rings}

The {\it collective} density fluctuations dynamics
is embedded in the {\it coherent intermediate scattering function}
in general defined as 
\begin{equation}
F_t(q,t)=\frac{1}{NS(q)}\langle\sum_{i=1}^{N}\sum_{j=1}^{N}e^{-i{\bar
q}\cdot \left[{\bar x}_i(t)-{\bar x}_j(0)\right]}\rangle
\label{def_fqt}
\end{equation}
where $N$ is the number of molecules involved and $S(q)$ the static
structure factor. In the present case the position variables $x_k(t)$ can be
identified with different quantities; here we are interested in
the dynamics of the molecular centers of mass and of the phenyl rings
so that we will consider the following scattering functions

\begin{eqnarray}
F_{t}^{(M)}(q,t)&=&\frac{1}{N_{M}S^{(M)}(q)}\langle\sum_{\xi'\xi''}e^{-i{\bar q}\cdot\left[
{\bar M}_{\xi'}(t)-{\bar M}_{\xi''}(0)\right]}\rangle \\
F_{t}^{(R)}(q,t)&=&\frac{1}{N_{R}S^{(R)}(q)}\langle\sum_{ij}\sum_{\xi'\xi''}
e^{-i{\bar q}\cdot\left[
{\bar R}_{i\xi'}(t)-{\bar R}_{j\xi''}(0)\right]}\rangle
\end{eqnarray}
Here ${\bar M}_{\xi'}(t)$ is the position of the center of mass
of the molecule $\xi'$ at time $t$ ($\xi'=1 \ldots N_M$), ${\bar R}_{i\xi'}(t)$
the position of the center of mass of the phenyl ring $i$ ($i=1 \ldots 3$) 
pertaining to the molecule  $\xi'$;
the functions are renormalized to the corresponding 
static structure factors. From now on, the superscripts
(R) and (M) will be referred to molecular and rings quantities
respectively.

As in the case of the incoherent scattering function,
at every temperature investigated we have reconstructed 
the whole curve, even on very short time scales,
by mean of two sets of system configuration
campionated with different frequencies (see Table III of Ref.~\cite{modilruosam}).
At every investigated temperature, we considered the momentum 
values $q_1=14$ nm$^{-1}$ and $q_2=19$ nm$^{-1}$
corresponding to the first and second peak of the experimental static 
structure factor, averaging on the values of $q$ falling
in the interval $q\pm \Delta q$ with $\Delta q=0.2$ nm$^{-1}$.
Moreover, the momentum dependence of the principal dynamical parameters
has been investigated at $T=280, 300, 330$ K for values of momenta
ranging from $2$ to $30$ nm$^{-1}$.

We made a long time analysis in term of the usual stretched
exponential form Eq.~(\ref{stretched}) determining the temperature and momentum
dependence of the fitting parameters $\tau_\alpha$, $\beta_\alpha$
and $f_q$ and verifying the TTSP.

\begin{figure}
\hbox to\hsize{\epsfxsize=1.0\hsize\hfil\epsfbox{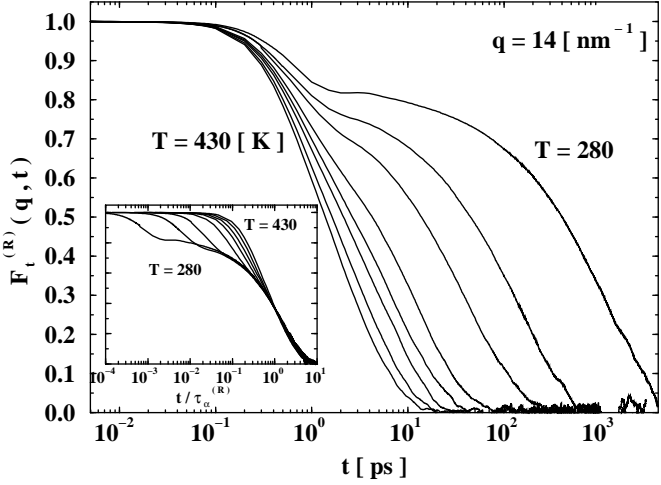}\hfil}
\vspace{.4cm}
\caption{Intermediate coherent
scattering functions $F_t^{(R)}(q,t)$ calculated 
on phenyl rings at $q=14$ nm$^{-1}$
for the temperatures $T=280, 300, 320, 350, 370, 390, 410, 430$ K
(from top to bottom);
in the inset we show the same curves rescaled
as a function of $t/\tau_{\alpha}^{(R)}$.}
\label{fqt}
\end{figure}

\begin{figure}
\hbox to\hsize{\epsfxsize=1.0\hsize\hfil\epsfbox{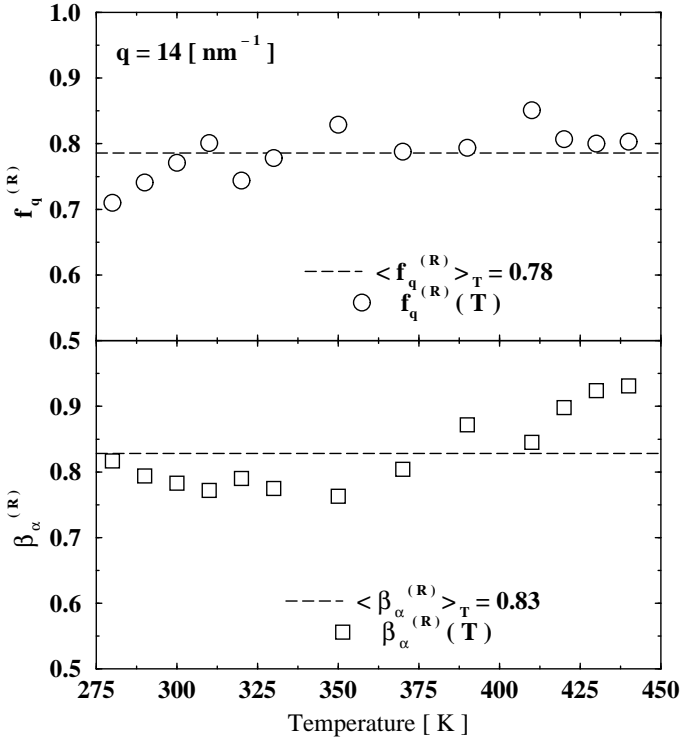}\hfil}
\caption{Temperature dependence of the stretched exponential parameters
calculated from $F_t^{(R)}(q,t)$
together with the corresponding mean values (dashed lines).
Top: non-ergodicity parameter $f_q^{(R)}(T)$. Bottom:
stretching parameter $\beta_\alpha^{(R)}(T)$.
}
\label{stretch}
\end{figure}

In Fig.~\ref{fqt} we show $F_t^{(R)}(q,t)$ 
calculated at $q_1$
for the temperatures $T=280, 300,320,350,370,390,410,430$ K (from top to
bottom); as in the case of the 
self dynamics, every curve decays to zero in the considered
time window and it is clearly visible the two steps decaying pattern.
The long time part of these $F_t^(R)$ have been fitted to
Eq.~(\ref{stretched}) and the parameters $\tau_\alpha$, $\beta_\alpha$ and $f_q$
are determined by a least square fitting routine. 
In the inset we plot the same curves as a function
of the rescaled time ${\bar t}=t/\tau_\alpha^{(R)}$; all the curves
collapse pretty well on a single
master curve as predicted by the TTSP.
The temperature dependence of the non-ergodicity parameter $f_q^{(R)}$
(top panel)
and of the stretching parameter $\beta_\alpha^{(R)}$ (bottom panel)
are shown in Fig.~\ref{stretch}; they are 
temperature independent, as predicted by MCT.
The mean values $f_q^{(R)}=0.78$ and $\beta_\alpha^{(R)}=0.83$ (dashed line) 
have to be compared with the values determined in the case of the 
self dynamics $f_q\simeq 0.7$ and
$\beta_\alpha\simeq0.8$~\cite{modilruosam}.
The two values of $\beta_\alpha$ are equal in the limit of the
error bars; at variance, the value for $f_q^{(R)}$ in the collective 
case is greater than the  value found in the one particle case.

In Fig.~\ref{tau_T_mol_rng} we plot the structural relaxation 
times at $q_1$ and $q_2$
for molecules (triangles and diamonds) and phenyl rings (circles and squares) 
in order to test if both $\tau_\alpha^{(M)}$ and $\tau_\alpha^{(R)}$
follow the same power law which is supposed to be momentum independent.
Both sets of data have found to be consistent with a power law
of the form Eq.~(\ref{powerlaw}) with parameters
$T_c\simeq 268$, $\gamma\simeq 2.3$;
these values have to be compared with the results
concerning the one particle dynamics $T_c=276\pm7$ K
and $\gamma=2.0\pm0.4$~\cite{modilruosam}. 
The inset show the data plotted as a function of ${\bar T}=T-T_c$,
in order to stress the power law dependence, and rescaled by an arbitrary
factor in order to maximize the overlap.

We now consider the momentum dependence of the collective dynamics
at few selected temperatures $T=280, 300, 330$ K, spanning the momentum region  
in the interval $2\div 30$ nm$^{-1}$. We test the long time 
dynamics in term of the stretched
exponential function and we verify the MCT predictions
on the von Schweidler time region, characterized by the power exponent $b$.

In Fig.~\ref{stretch_q} (top panel) we show the stretched exponential fits (solid lines)
to $F_t^{(R)}$ at some selected values of $q$; they works
pretty good at least for time values greater than $5$ picoseconds
as is more clear looking at the bottom panel. Here we show the whole calculated 
curve (circles) up to $t\simeq 0.05$ ps at $q=14$ nm$^{-1}$ and $T=300$ K
together with the stretched exponential fit (solid line).
From the top panel of Fig.~\ref{stretch_q} is qualitatively clear 
that the relaxation time depends non trivially on the momentum values.
As we reminded in Sect.~\ref{MCT}, MCT predicts that in the 
collective case, at fixed temperature, the relaxation times oscillate in phase
with the static structure factor, i. e. they are strongly coupled
to the static structure of the system.

In Fig.~\ref{tau_q_mol} and Fig.~\ref{tau_q_rng}
we plot the momentum dependence of the collective relaxation times (top panels)
for molecules and rings respectively, at $T=280, 300,330$ K 
(circles, squares and triangles respectively) together with the 
corresponding static structure factors divided by $q^2$ (bottom panels). 
The correlation among these two quantities, predicted by 
Eq.~(\ref{tau_resc}), is evident; it is also evident
that, at variance with the temperature dependence, the
momentum dependence of the relaxation times is completely
different in the two cases.

In the molecular case (Fig.~\ref{tau_q_mol}) only a maximum at 
$q\simeq 9$ nm$^{-1}$ corresponding to intermolecular correlations
is present ( a small shoulder at $q\simeq 14$ nm$^{-1}$,
related to correlations between rings pertaining to different molecules,
can be also identified).
In the case of the phenyl rings (Fig.~\ref{tau_q_rng})
the momentum dependence is much more structured and three 
main features are present:

(i) a maximum at $q\simeq 9$ nm$^{-1}$ related to 
correlations between molecular centers of mass

(ii) a shoulder at $q\simeq 14$ nm$^{-1}$ 
(well developed in a maximum at the lowest temperature $T=280$ K)
related to correlations between rings belonging to different molecules

(iii) a maximum at $q\simeq 22$ nm$^{-1}$ related to
correlations between rings pertaining to the same molecules;
this is the more important features
related mainly to the orientation of the
lateral rings with respect to the central ring.

It is clear from this result that in complex molecular
glass formers there are intramolecular rotational and vibrational
degrees of freedom that couple to the translational long-time modes
as already stressed in Ref.~\cite{tollschowutt}; a theory 
not taking into account these degrees of freedom
cannot explain the whole momentum dependence of the 
centers of mass dynamics.

In Fig.~\ref{tau_q_resc} we show the different relaxation times
(left panel for molecules and right panel for rings) at the three 
selected temperatures
multiplied by the corresponding diffusion
coefficients from Ref.~\cite{modilruosam}; 
as predicted by Eq.~(\ref{tau_resc}),
these products are temperature independent
for values of $q$ close to the first maximum of the static
structure factor.

In Fig.~\ref{stretch_q_rng}(A) we plot the momentum dependence
of the stretching parameter $\beta_\alpha^{(R)}$
while in  Fig.~\ref{stretch_q_rng}(B)
we show the non ergodicity parameter $f_q^{(R)}$ at the three selected
temperatures together with $S^{(R)}(q)/q^2$ ( Fig.~\ref{stretch_q_rng}(C));
also in this case a strong correlation between the
oscillations of the quantities is clear, as expected.

We have seen in Sec.~\ref{MCT} that the long time limit 
of the $\beta$-region can be described by the von Schweidler power law 
Eqs.~(\ref{betaregion})~(\ref{bpower}).
The exponent $b$ is expected to be momentum independent
and to assume the same value of the self dynamics case
namely $b=0.52$~\cite{petbarpuj}; 
at the contrary, the amplitude $h(q)$ is expected
to be momentum dependent and to oscillate out of phase with
the static structure factor.
We then calculated a power law fit in the form 
$F_t^{(R)}(q,t)=f_q^{(R)}-c_2^{(R)}(q)\,t^{b^{(R)}}$
for all values of momentum considered, in a time window depending on
the particular q-value but always included in the interval
$2\div 30$ ps; moreover we considered the three 
parameters free as in the case of the self dynamics.
All the observations done in the previous work concerning 
the great uncertainties on the estimated values of
the fitting parameters, hold in the present case.
In Fig.~\ref{beta_q_rng}
(A) we plot the power exponent $b(q)$ that is supposed
to be momentum independent; some smooth oscillations are
nevertheless present but this can be due
to the interplay during the fitting procedure
with the other oscillating parameters.
In Fig.~\ref{beta_q_rng}(B) we plot $f_q^{(R)}$;
these points must be compared with the results 
of Fig.~\ref{stretch_q_rng} (B) and the agreement looks quite nice
as expected. The value of the plateau, indeed,
must be the same if determined as the small time limit
of the $\alpha$-region or the long time limit of the $\beta$-process.
In Fig.~\ref{stretch_q_rng} (C)
we plot the quantity $1/c_2^{(R)}(q)$ and also in this case
oscillations in phase with $S^R(q)/q^2$ are evident,
as expected.

\section{NEUTRON SCATTERING}
\label{neutrons}

Neutron scattering is
one of the most powerful tools used in the study of supercooled liquids
and glasses in the $q-$region covered by the MD simulations. 
Experimentally the scattering function $F_{t}(q,t)$ of
Eq.~(\ref{def_fqt}) can be determined~\cite{bartfujleg}
by neutron scattering experiments
either directly, on neutron spin echo instruments, or Fourier transforming
the dynamical structure factor $S(q,\omega)$

\begin{equation}
\frac{S(q,\omega)}{S(q)}=\frac{1}{2\pi\hbar}\int dt e^{-i\omega t} F_{t}(q,t) 
\end{equation}
calculated by means of triple axis backscattering or time of flight spectroscopy.
The experimental neutron scattering cross section $(d\sigma/d \Omega
dE)$ is generally composed of a {\em coherent} and an {\em incoherent} part

\begin{eqnarray}
d\sigma/d\Omega dE&\simeq &<b>^2 S_{coh}(q,\omega)\\
&+&[<b^2>-<b>^2] S_{incoh}(q,\omega)\nonumber
\end{eqnarray}
where $b$ is the scattering length and the symbol
$\langle\rangle$ denotes an average over 
the distribution of nuclear spins and isotopes.
The isotopic composition of the sample
allows us to study selectively the collective motion
via coherent scattering from deuterated samples
(the scattering lengths of D and C atoms are basically coincident)  
and the one particle motion via
incoherent scattering from protonated samples.

The interaction of neutrons with a bulk sample of OTP
can be simulated numerically taking into account 
the interactions of neutrons
with both carbon (C) and deuterium (D) atoms.
H atoms are not considered in our dynamics but is a
reasonable approximation to put them in fixed position
on the line extending from the center of the ring
trough a carbon atom at the fixed C-H distance $d_{C-H}=0.107$ nm;
so that, knowing the coordinates of the rings, is trivial 
to reconstruct their own positions.
We then define a neutron (N) coherent scattering function
$F_t^{(N)}(q,t)$ as

\begin{eqnarray}
F_{t}^{(N)}(q,t)&=&\frac{1}{N_{A} S^{(N)}(q)}\langle\sum_{\lambda'\lambda''}\sum_{ij}\sum_{\xi'\xi''}\\
&&
b_{\lambda'}b_{\lambda''}e^{-i{\bar q}\cdot\left[
{\bar r}_{\lambda'i\xi'}(t)-{\bar r}_{\lambda''j\xi''}(0)\right]}\rangle \nonumber
\end{eqnarray}
where $N_A=N_C+N_H$ (3456 in this case) is the total number of atoms, ${\bar r}_{\lambda'i\xi'}(t)$
is the position of the atom $\lambda'$ pertaining to the ring $i$
in the molecule $\xi'$ and $S^{(N)}(q)$ is the static structure factor
of Fig.~11 of Ref.~\cite{modilruosam}. The number of hydrogen atoms
is $4$ for each central ring and $5$ for each lateral ring.
The scattering lengths $b_{\lambda}$ are in principle 
different for the carbon and deuterium atoms
but, as we observed in Ref.~\cite{modilruosam},
they are both positive and of the same magnitude so that
is a good approximation to consider
the product $b_{\lambda'}b_{\lambda''}$ an ineffective
positive constant.
The function $F_{t}^{(N)}(q,t)$ is the quantity directly
comparable with the experimental data.

In the present section we present a comparison
between the temperature and momentum dependencies 
of the MD and experimental spectra of Ref.~\cite{tollwutt}
calculated from perdeuterated $C_{18}D_{14}$ by means of
coherent neutron time-of-flight and backscattering spectroscopy.

At this stage few remarks must be done about the 
momentum dependence of the MD and experimental sets of data.
All these data are supposed to 
depend on the structure of the systems so that, in general,
some differences are expected
(experimental and MD structures are slightly different~\cite{modilruosam}).
Anyway, two observations about the experimental results 
must be made. First of all, as reported in Ref.~\cite{tollwutt}, 
some reservation is necessary for the 
experimental data at the smallest 
momenta, $q\leq 6$ nm$^{-1}$, where $f_q$ tends
toward $1$. Indeed, in this region one expects
significant background from incoherent scattering, which contributes
about $15\%$ of the total cross section, 
and from multiple scattering.
Moreover the technique used to determine the values of the dynamical
parameters from the experimental data are quite different 
with respect the MD computation.
Indeed, due to the limited dynamical window of the available
spectrometers,  in the experimental case  a direct
fit of the data to the stretched exponential function
with three independent parameters is
not possible. Based on the observations that $\tau(q)\propto \eta(T)/T$
($\eta(T)$ is the viscosity at temperature $T$)
and that the line shape is independent on temperature,
at fixed momentum $q$ the spectra
at different temperatures are rescaled in time to
$\bar t=t/t_s$ where the scaling time is given by $t_s=\tau_\eta
(T)/\tau_\eta (T=290 K)$ and $\tau_\eta=\eta(T)/T$. In this way the data
converge toward a temperature-independent long time asymptote;
this is the curve actually fitted to the stretched exponential.

In Fig.~\ref{compar_T} we show the temperature dependence
of the stretched exponential fitting parameters for the $\alpha$-region
of both MD (open circles) and experimental (solid circles) data
determined at $q=14$ nm$^{-1}$.
In panel (A) we show the structural relaxation times; the inset shows
the same data but the MD points have been shifted of $20$ K
as already explained in Ref.~\cite{modilruosam}.
We stressed out, indeed, that a non perfect 
parameterization of the diffusive behavior of the model system
controlled by the value of $\epsilon_{LJ}$, has shifted the MD thermodynamical
point about $20$ K above the corresponding experimental temperature.
From the figure is clear that the agreement among the two sets of data
is very good on a very wide time region.
In panel (B) we show our results for the non ergodicity parameter $f_q$
and also in this case the agreement is very good among the two sets of
data; in panel (C) we plot the stretching parameters $\beta_{\alpha}$. In this
case the MD points are systematically above the experimentally
determined data. This effect can be due to the observations
made before; moreover, for the determination of the $\beta_{\alpha}$ parameter, 
the very long time points are crucial and they seem to lack
in the experimental data analysis.
 
With Fig.~\ref{neutron_q_MD} we start the comparison of the momentum dependence
of the two sets of data. 
In the top panel of Fig.~\ref{neutron_q_MD} we plot the simulated scattering functions
$F_t^{(N)}(q,t)$ (symbols) at $T=290$ K at the indicated $q$
together with the stretched exponential long time fits (solid lines).
In this case $f_q^{(N)}$ is expected to be momentum dependent so
that, in order to have a master plot (bottom panel of Fig.~\ref{neutron_q_MD}),
we have to show ${\bar F}_t^{(N)}(q,t)=F_t^{(N)}(q,t)/f_q^{(N)}$ as a function of
the rescaled time ${\bar t}=t/\tau_\alpha^{(N)}$. The rescaled curves 
collapse quite well on a master curve as expected.
In the next three figures we show the comparison 
for the momentum dependence of the MD data
at the three temperatures $T=280, 290, 300$ K
and the experimental data at $T=313, 320,330$ K;
the two sets of temperatures should be comparable due to
the $20$ K shift of the MD data. 

In Fig.~\ref{tau_q_compar} we show the 
MD (left panel) and experimental (right panel) relaxation times
together with the corresponding static structure factors
renormalized to $q^2$. Striking similarities are clear
even if maxima on MD results correspond to bumps of experimental
results. The clearest difference is the decrease of
$\tau_\alpha^{(N)}$ at small value of $q$ that appear to be
completely absent in the experimental case.
This could be due to the incoherent background on the
experimental data stressed above; moreover, 
is well known that sometimes MD is not able to
determine the correct value of $\tau$ at small value of $q$
but usually it tends to overestimate its correct value,
at variance of this case.
Anyway, the agreement among the two sets of data at the higher
values of $q$ corresponding to the intramolecular correlations
is surprising; none rigid model could be able to do that.

In Fig.~\ref{f_q_compar} we show the results concerning the non
ergodicity parameter $f_q^{(N)}$. The different curves are temperature 
independent as expected;
both sets of data show a maximum at $q\simeq14$ nm$^{-1}$, while
the maximum at $q\simeq 9$ nm$^{-1}$ for the MD data correspond to a
little shoulder in the experimental case. Also in this case a decreasing 
part at small $q$ is present in the MD data at variance with the
experimental results. But according to Fig.~7 of Ref.~\cite{tollschowutt}
the calculation of the non-ergodicity parameter fitting the $\beta-$
region gives a plateau at $q\simeq 6$ nm$^{-1}$ at variance
with the increasing behavior of the $\alpha-$ region analysis.
However, as already stressed in Sec.~\ref{mol_rings}, these data 
are supposed to coincide. Concluding, also in this case,
the experimental data at small $q$ seem to be not reliable.

In Fig.~\ref{beta_q_compar} we finally show the stretching parameter
$\beta_\alpha^{(N)}$. MD data are very noisy,
nevertheless it is possible to recognize an oscillatory behavior; 
moreover, the points at the smallest available
value of $q$ seem to catch the decreasing behavior of the
experimental data.

\section{SUMMARY AND CONCLUSIONS}
\label{conclusions}

In this paper we have concluded the analysis of the long time
center of mass dynamics of the intramolecular model for OTP
introduced in Ref.~\cite{modilruosam}; there we found 
good agreement among the diffusion properties
of the simulated and the real system as well as among the two
single-particle dynamics. Moreover in Ref.~\cite{modilruosam}
we found a very good agreement
with the main predictions of MCT. In the present paper
we definitely confirm such an agreement.

We have studied the collective density fluctuations 
on a large temperature and momentum range range,
considering both the fluctuations related to molecular and phenyl ring
centers of mass. With respect to the temperature dependence we 
found the usual double step decaying pattern and we confirmed
the main predictions of the MCT about the behavior of
the stretched exponential parameters; in particular the 
relaxation times obey the same power law.
The momentum dependence of the stretching parameters appears
really interesting; MCT predicts that the behavior in
the momentum space of the structural relaxation
time, at variance with the the trivial square 
law of the one-particle case, is driven by the structure,
namely it is proportional to $S(q)/q^2$. 
In both cases, molecules and phenyl rings, this prediction
is completely fulfilled. The phenyl rings 
behavior is particularly interesting;
every feature of the quantity $S(q)/q^2$ is mirrored 
on the $\tau_\alpha^{(R)}(q)$ curve at the three investigated
temperatures $T=280, 300,330$ K up to a 
value of momentum $q\simeq 30$ nm$^{-1}$.
In particular, the maximum at $q\simeq 22$ nm$^{-1}$
is related to fluctuations taking place on molecular length scales;
the long time structural dynamics appears, indeed, coupled to the
dynamics of internal molecular degrees of freedom.
Similar oscillations are present for the other
stretched parameters and also the correct behavior for
the short time $\beta$-region is found.

The next step has been a comparison among the experimental
and simulated neutron scattering spectra calculated considering
both the scattering from carbon and deuterium atoms.
The temperature dependences of the relaxation times in both cases 
are well described by the same power law
for almost three decades in time; a good agreement is also found for the
other stretching parameters.
The momentum dependence is qualitatively very similar
in both cases; although some different features not clear are present.  

\begin{figure}
\hbox to\hsize{\epsfxsize=1.0\hsize\hfil\epsfbox{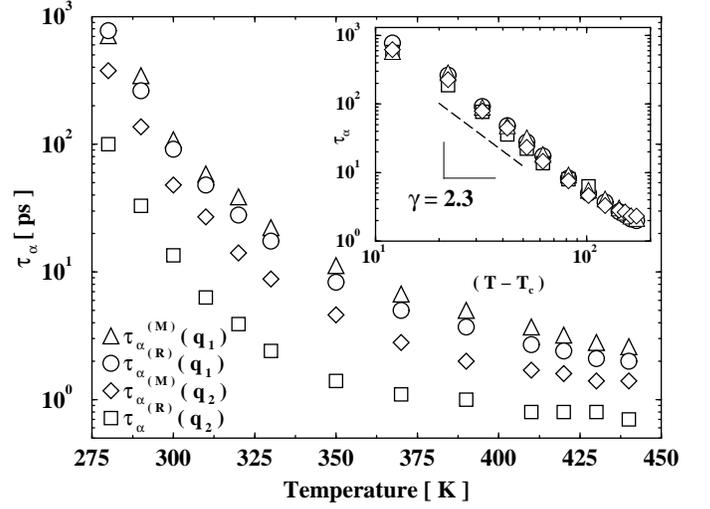}\hfil}
\vspace{1.cm}
\caption{Temperature dependence of the structural relaxation times
at $q_1=14$ nm$^{-1}$ and $q_2=19$ nm$^{-1}$ calculated both on
rings (circles and squares respectively) and
molecules (triangles and diamonds respectively) centers of mass.
In the inset the date are shown in a double-log scale as a function of
the rescaled temperature $(T-T_c)$; the points have been shifted in
order to maximize the mutual overlap and to stress
the power law behavior.
The power law of exponent $\gamma=2.3$ is also shown (dashed line);
the value for $T_c$ is $268$ K.}
\label{tau_T_mol_rng}
\end{figure}

\begin{figure}
\hbox to\hsize{\epsfxsize=1.0\hsize\hfil\epsfbox{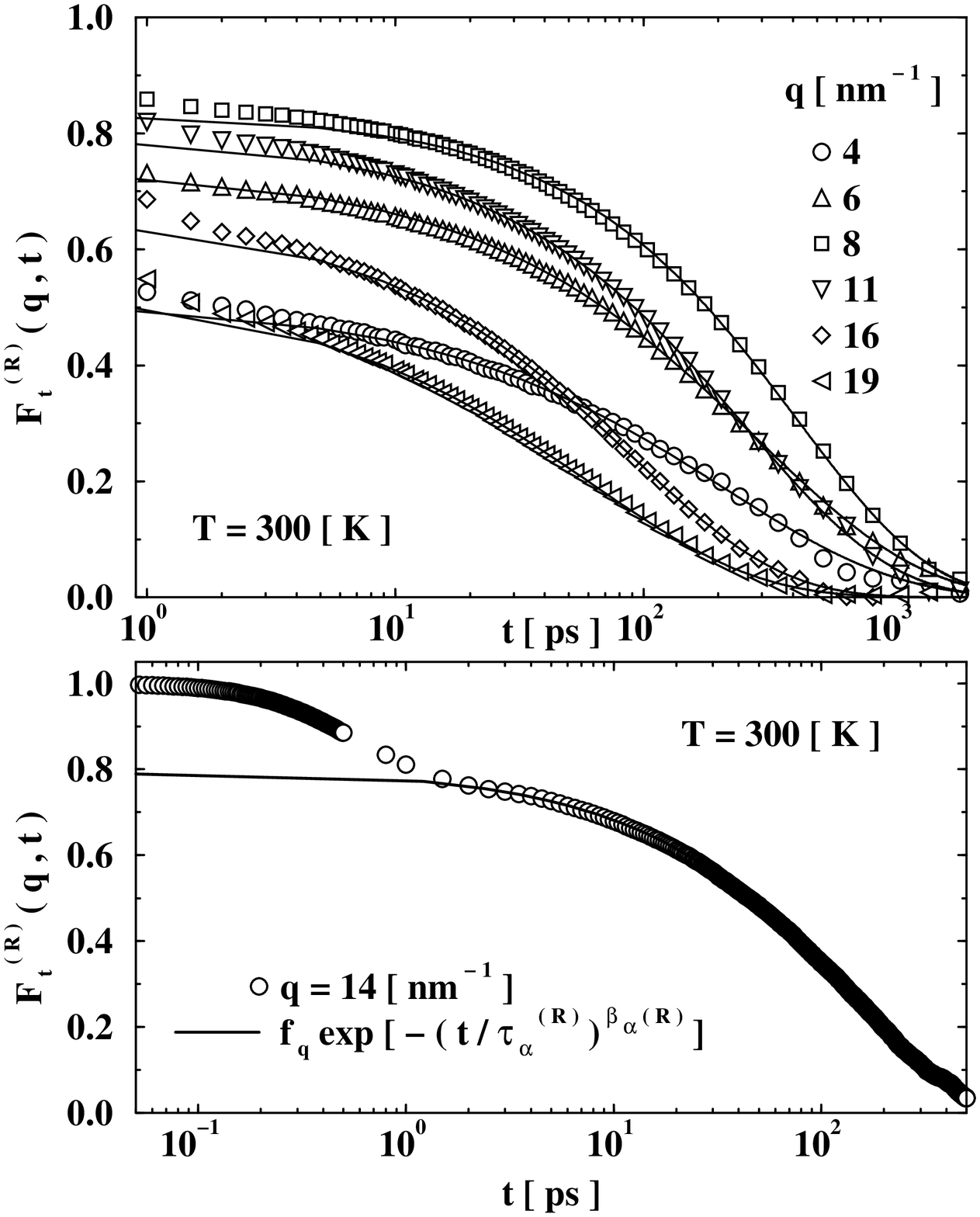}\hfil}
\caption{Top: Intermediate coherent scattering functions $F_t^{(R)}(q,t)$ calculated
at fixed temperature $T=300$ K for different values of momentum $q$;
the corresponding long-time stretched exponential fits are also shown (solid lines).
Bottom: Here we show the whole time behavior of $F_t^{(R)}(q,t)$ at
$q=14$ nm$^{-1}$ together with the stretched exponential fit. 
}
\label{stretch_q}
\end{figure}

\begin{figure}
\hbox to\hsize{\epsfxsize=1.0\hsize\hfil\epsfbox{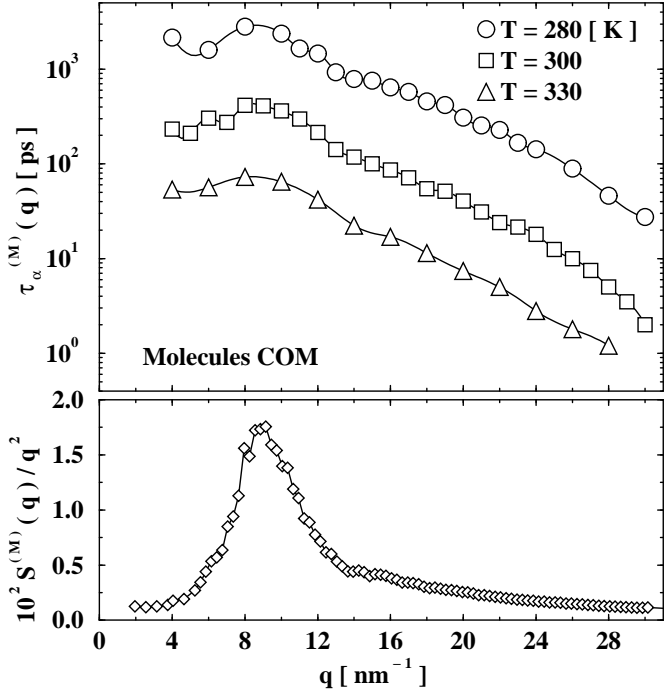}\hfil}
\caption{De Gennes narrowing. Top: Momentum dependence of the structural relaxation time
calculated considering molecular centers of mass at the three
temperatures $T=280, 300, 330$ K (circles, squares and triangles
respectively). Bottom: Structure factor calculated on molecular
centers of masses divided by $q^2$.}
\label{tau_q_mol}
\end{figure}

\begin{figure}
\hbox to\hsize{\epsfxsize=1.0\hsize\hfil\epsfbox{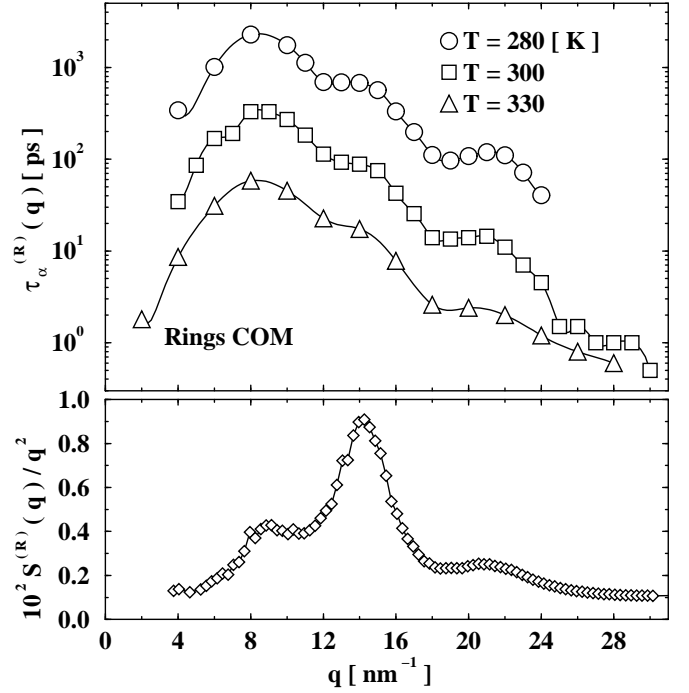}\hfil}
\caption{Top: Momentum dependence of the structural relaxation time
calculated considering phenyl rings centers of mass at the three
temperatures $T=280, 300, 330$ K (circles, squares and triangles
respectively). Bottom: Structure factor calculated on rings
centers of masses divided by $q^2$.}
\label{tau_q_rng}
\end{figure}

\begin{figure}
\hbox to\hsize{\epsfxsize=1.0\hsize\hfil\epsfbox{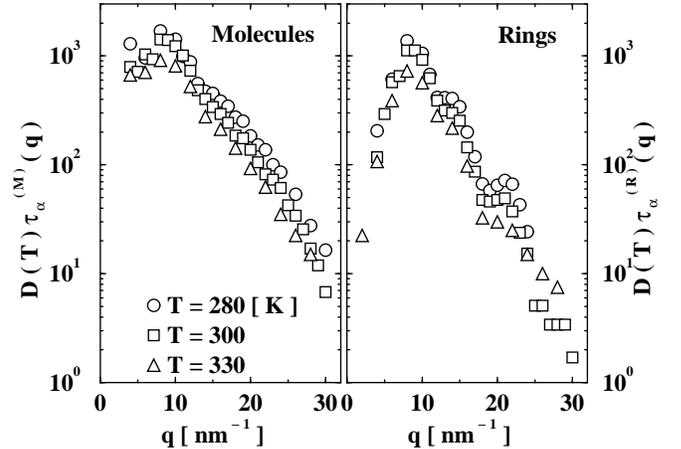}\hfil}
\vspace{0.4cm}
\caption{Left: Structural relaxation times $\tau_\alpha^{(M)}$ for molecules
at temperatures $T=280, 300, 330$ K multiplied by the correspondent 
diffusion coefficients $D(T)$; this product is supposed to be temperature
independent. Right: as the left panel but
for $\tau_\alpha^{(R)}$
calculated on phenyl rings centers of masses.}
\label{tau_q_resc}
\end{figure}

\begin{figure}
\hbox to\hsize{\epsfxsize=1.0\hsize\hfil\epsfbox{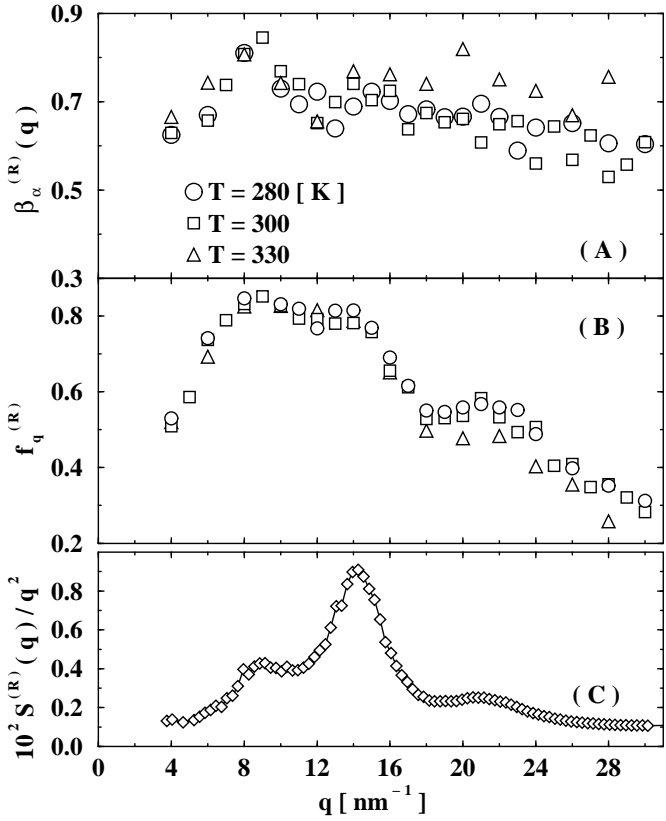}\hfil}
\caption{Momentum dependence of the stretching parameters $\beta_\alpha^{(R)}$
(A) and $f_q^{(R)}$ (B); oscillations
in phase with the structure factor (C) are evident.}
\label{stretch_q_rng}
\end{figure}

\begin{figure}
\hbox to\hsize{\epsfxsize=1.0\hsize\hfil\epsfbox{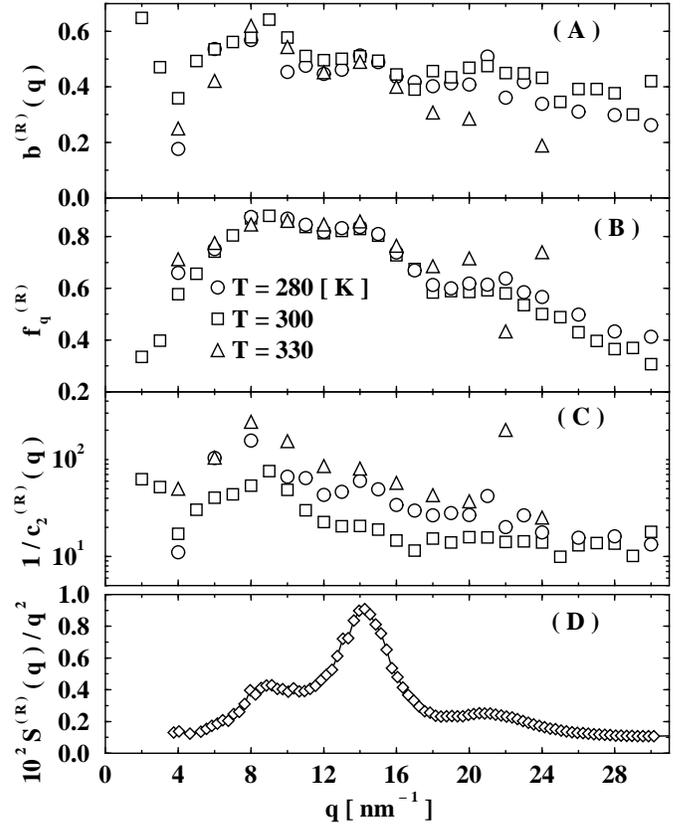}\hfil}
\caption{Momentum dependence of the fitting parameters for the
$\beta$-region
$b^{(R)}(q)$ (A), $f_q^{(R)}$ (B) and $1/c_2^{(R)}(q)$ (C); also in this case
a clear correlation with the static structure factor (D) is evident.}
\label{beta_q_rng}
\end{figure}

\begin{figure}
\hbox to\hsize{\epsfxsize=1.0\hsize\hfil\epsfbox{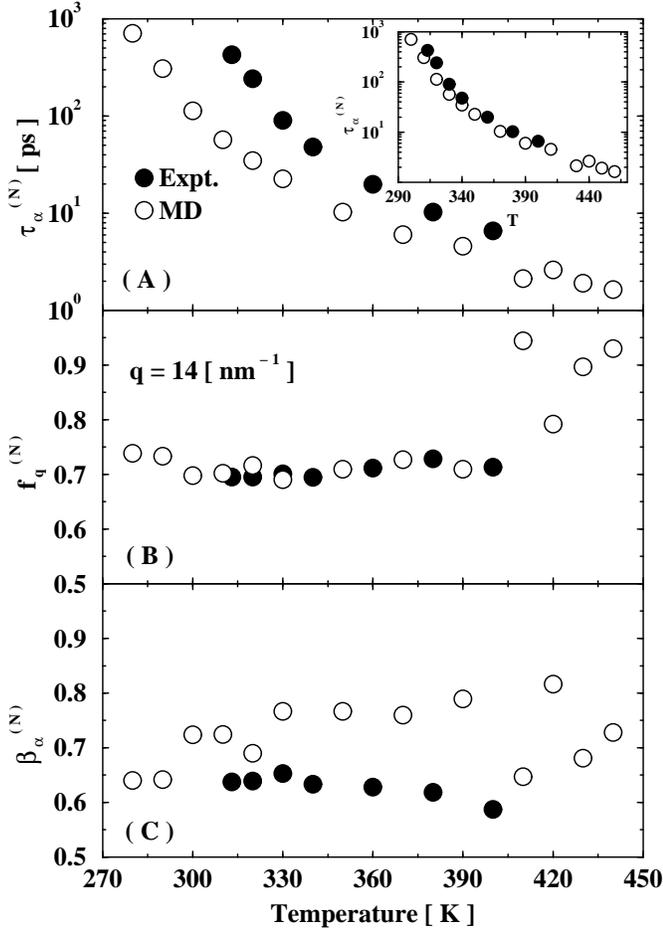}\hfil}
\caption{Comparison among the temperature dependencies of the 
stretched exponential parameters as calculated by MD
simulated neutron spectra (open symbols) and experimental neutron
scattering (full symbols) at $q=14$
nm$^{-1}$. (A) relaxation time $\tau_\alpha^{(N)}$; in the inset the MD data have been
shifted of $20$ K as explained in the text. (B) non ergodicity parameter $f_q^{(N)}$. (C) stretching
parameter $\beta_\alpha^{(N)}$.
}
\label{compar_T}
\end{figure}

\begin{figure}
\hbox to\hsize{\epsfxsize=1.0\hsize\hfil\epsfbox{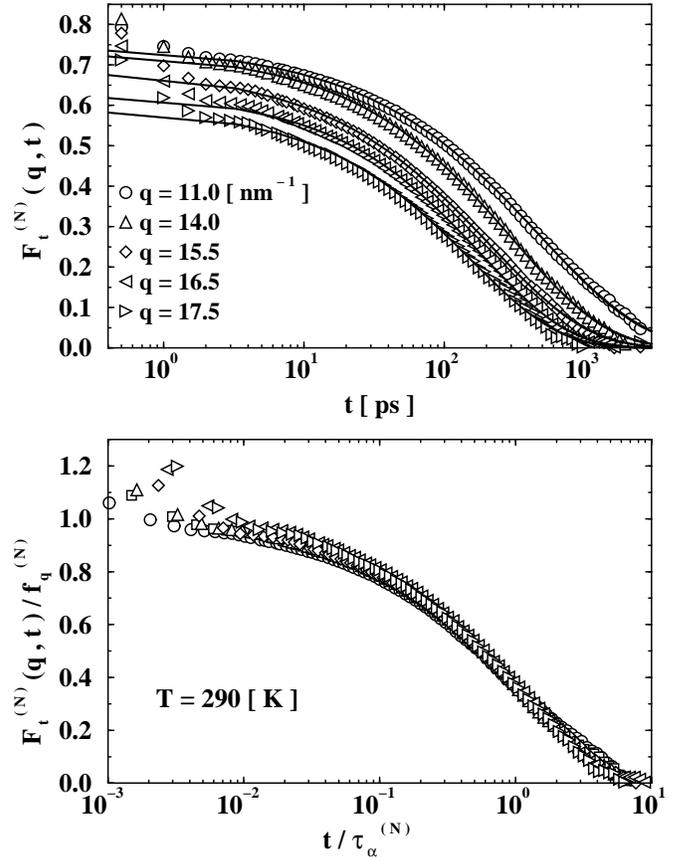}\hfil}
\caption{{\em Top}: Simulated neutron coherent scattering function $F_t^{(N)}(q,t)$ 
at $T=290$ K as a function of momentum. {\em Bottom}: $F_t^{(N)}(q,t)/f_q^{(N)}$
plotted
as a function of the rescaled time $t/\tau_\alpha^{(N)}$. All the curves collapse
on a master curve as expected (in this case $f_q^{(N)}$ depends on $q$).}
\label{neutron_q_MD}
\end{figure}

\begin{figure}
\hbox to\hsize{\epsfxsize=1.0\hsize\hfil\epsfbox{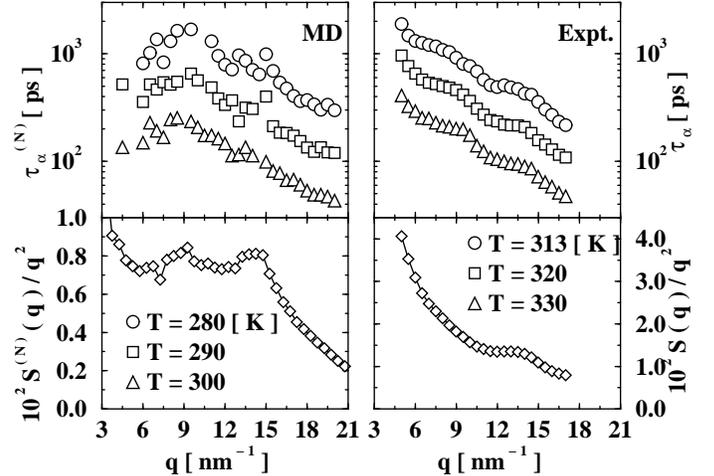}\hfil}
\vspace{0.4cm}
\caption{Comparison between the values of $\tau_\alpha^{(N)}$ calculated by MD
simulations (left panel)
and the experimental ones (right panel) together with the appropriate structure
factors divided by $q^2$.}
\label{tau_q_compar}
\end{figure}

\begin{figure}
\hbox to\hsize{\epsfxsize=1.0\hsize\hfil\epsfbox{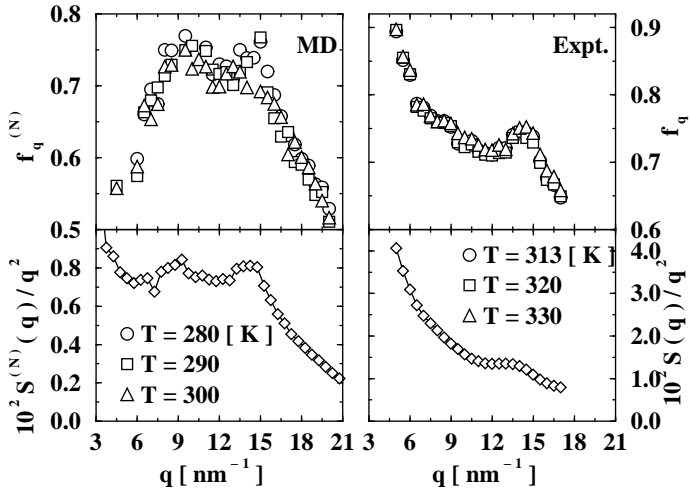}\hfil}
\vspace{0.4cm}
\caption{As above for the non-ergodicity parameter $f_q^{(N)}$.}
\label{f_q_compar}
\end{figure}

\begin{figure}
\hbox to\hsize{\epsfxsize=1.0\hsize\hfil\epsfbox{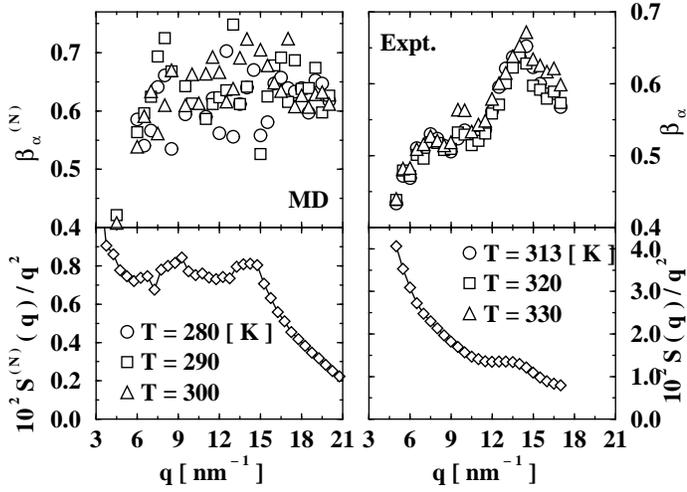}\hfil}
\vspace{.4cm}
\caption{As above for the stretching factor $\beta_\alpha^{(N)}$.}
\label{beta_q_compar}
\end{figure}

\begin{figure}
\hbox to\hsize{\epsfxsize=1.0\hsize\hfil\epsfbox{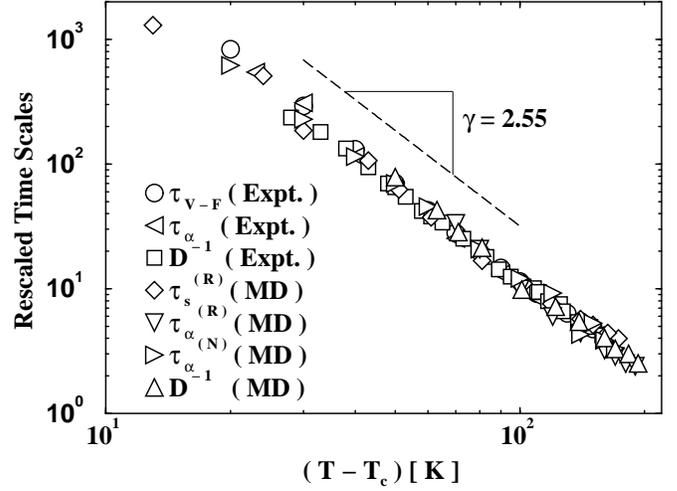}\hfil}
\caption{Master plot of all the time scales related to the center of mass
dynamics of the system, calculated by means of molecular dynamics and measured
experimentally: experimental viscosity (circles), 
neutron scattering collective experimental relaxation time (left
triangles),
inverse of the experimental diffusion coefficient (squares),
MD one particle
(diamonds) and structural relaxation time (triangles
down) calculated on phenyl rings,  
MD neutron spectra relaxation time (right triangles),
inverse of the MD diffusion coefficients (triangles up). 
The data have been rescaled by arbitrary constants in order
to maximize the overlap with the viscosity experimental results. 
All these time scales follow the same MCT 
power law with exponents $T_c=290$ K and $\gamma=2.55$.}
\label{tau_glob_final}
\end{figure}

At this point some conclusions can be drawn
about the capability of our model to describe
the long time dynamics of the real system.
All the center of mass time scales calculated by molecular dynamics
have been found to be consistent with a power law,
although some discrepancies were present for the actual value 
of the power exponent mainly due to uncertainties
on the fitting procedure. Moreover it has been shown
that our actual MD thermodynamical point is
shifted by about $20$ K with respect to the corresponding
experimental point. 
Taking into account all these informations, we plot in
Fig.~\ref{tau_glob_final} all the time scales
related to the centers of mass dynamics
considered up to now, both experimental and numerical,
as a function of the rescaled temperature $\tilde{T}=T-T_c$
where $T_c=290$ K~\cite{petbarpuj} for the experimental points
and $T_c=270$ K for the MD results.
In particular,
$\tau_{V-F}(T)$ (circles) is the time scale related to the 
{\em shear viscosity} $\eta_s$
of Ref.~\cite{giuphd}, 
$\tau_{\alpha}(T)$ (left triangles) is the neutron scattering 
structural relaxation 
time of Ref.~\cite{tollwutt} at $q=14$ nm$^{-1}$ and the inverse of the 
experimental diffusion coefficient (squares)
is from Refs.~\cite{calldougfal}~\cite{fujgeilsill}.
The MD one-particle relaxation time calculated 
on phenyl rings $\tau_s^{(R)}(T)$
(diamonds) and self-diffusion coefficient (up triangles) 
are from Ref.~\cite{modilruosam}; finally, $\tau_\alpha^{(R)}(T)$ 
(down triangles) is the structural relaxation time of 
Fig.~\ref{tau_T_mol_rng} calculated at $q=14$ nm$^{-1}$
on phenyl rings and $\tau_\alpha^{(N)}(T)$ the simulated 
neutron scattering relaxation time of Fig.~\ref{compar_T}.
All the data collapse pretty well for almost 
three decades in time on the same line 
corresponding, on a double-log scale, to a power law of
the form of Eq.~(\ref{powerlaw}) of exponent $\gamma=2.55$~\cite{petbarpuj}.

Concluding, our model has found to be a very successful
model for the centers of mass dynamics of the real system,
showing a {\em critical} behavior 
consistent with the experimental results
in a wide time window. The implementation of an intramolecular dynamics
is relevant to such extent; in particular, the internal degrees
of freedom appear to be strongly coupled to the long time structural
dynamics as is also clear from the study of the rotational properties of the
system~\cite{phd}. 
The next step will be the understanding of the role played
by the internal degrees of freedom in the fast relaxations
observed experimentally~\cite{fastmd}. 

\begin{center}
{\bf ACKNOWLEDGMENTS}
\end{center}

The authors wish to thank F.~Sciortino for very useful discussions
and J.~Wuttke for the raw experimental data of Ref.~\cite{tollwutt}.

\newpage

\end{multicols}

\end{document}